\documentclass{article}
\usepackage{spconf,amsmath,graphicx}
\usepackage{booktabs}       % professional-quality tables
\usepackage{amsfonts}       % blackboard math symbols
\usepackage{amssymb}
\usepackage{multirow}
\usepackage{float}
\usepackage[dvipsnames]{xcolor}
\usepackage{colortbl}
\usepackage{bm}
\usepackage{arydshln}

\title{VIC-KD: Variance-Invariance-Covariance Knowledge Distillation to Make Keyword Spotting More Robust Against Adversarial Attacks}

\name{\begin{tabular}{c}Heitor R. Guimarães, Arthur Pimentel, Anderson Avila, Tiago H. Falk\end{tabular}}%
    
\address{%
Institut national de la recherche scientifique (INRS-EMT), Université du Québec, Montréal, Canada, and\\%
    INRS-UQO Mixed Research Unit on Cybersecurity, Gatineau,Québec, Canada
}
\begin{document}
%\ninept
%
\maketitle
\begin{abstract}
Keyword spotting (KWS) refers to the task of identifying a set of predefined words in audio streams. With the advances seen recently with deep neural networks, it has become a popular technology to activate and control small devices, such as voice assistants. Relying on such models for edge devices, however, can be challenging due to hardware constraints. Moreover, as adversarial attacks have increased against voice-based technologies, developing solutions robust to such attacks has become crucial. In this work, we propose VIC-KD, a robust distillation recipe for model compression and adversarial robustness. Using self-supervised speech representations, we show that imposing geometric priors to the latent representations of both Teacher and Student models leads to more robust target models. Experiments on the Google Speech Commands datasets show that the proposed methodology improves upon current state-of-the-art robust distillation methods, such as ARD and RSLAD, by 12\% and 8\% in robust accuracy, respectively.

%Keyword spotting (KWS) is an essential component for virtual assistants, with neural networks emerging as the primary choice for improved accuracy over traditional signal processing methods. However, designing KWS models for edge devices requires careful consideration of hardware constraints, and robustness against adversarial attacks has become a critical concern. In this work, we propose VIC-KD, a robust distillation recipe to tackle model compression and adversarial robustness jointly. Using self-supervised speech representation models as Teacher models, we show that imposing geometric priors to the latent representations of both Teacher and Student models can result in more robust target models. Experiments on the Google Speech Commands dataset show that the proposed methodology improves upon current state-of-the-art robust distillation methods, such as ARD and RSLAD, by 12\% and 8\% in robust accuracy, respectively.
\end{abstract}
\vspace{-1mm}
\section{Introduction}
\label{sec:intro}

\begin{figure*}[t]
        \centering
        \includegraphics[width=0.8\linewidth]{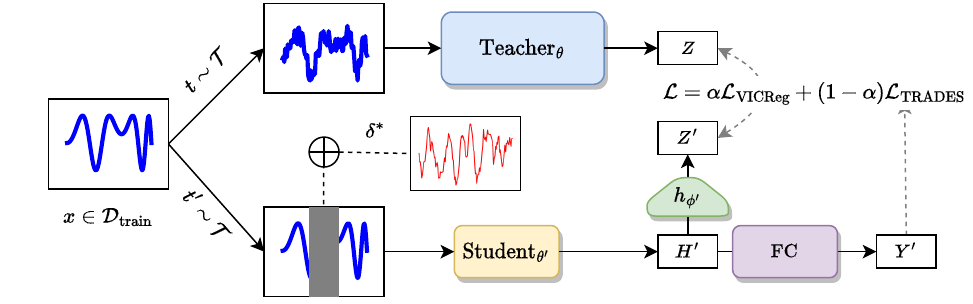} \vspace{-4mm}
        \caption{Diagram of our proposed method VIC-KD for robust distillation.}
        \label{fig:rgkd_diagram}
\end{figure*}

Recent advances in the deep learning field have profoundly impacted the speech-processing community. As such, it has resulted in cutting-edge systems for tasks such as keyword spotting (KWS)~\cite{chen2014small} and speech/emotion recognition~\cite{OSHAUGHNESSY2023101538, 8388669}, to name a few. In particular, KWS systems are usually the first layer of virtual assistants, where such models attempt to identify preset words in an utterance to activate the device (e.g., ``Hey Siri''). On-device speech processing, such as Google Now and Alexa, is becoming ubiquitous in our daily lives; however, this can impose challenges on energy efficiency, real-time processing, and on user privacy.

For KWS, the use of self-supervised speech representation learning (S3RL) has become a popular tool. S3RL models, such as \emph{Wav2Vec 2.0}~\cite{baevski2020wav2vec} and WavLM~\cite{chen2022wavlm}, are designed to learn combined acoustic and language models for continuous audio inputs. Notwithstanding, when working with edge devices, it is crucial to take into account model size due to hardware constraints. To this end, knowledge distillation (KD)~\cite{hinton_kd} has emerged as a powerful tool to transfer knowledge from a larger (Teacher) model to a smaller (Student) one, resulting in comparable generalization capabilities.

%When working with edge devices, it is crucial to consider the neural network architecture for optimal efficiency. Although self-supervised speech representation learning (S3RL) models can produce excellent results when fine-tuned for the KWS task, their size makes them unsuitable for deployment on edge devices. S3RL models, such as \emph{Wav2Vec 2.0}~\cite{baevski2020wav2vec}  and WavLM~\cite{chen2022wavlm}, are designed to learn combined acoustic and language models for continuous inputs. While it may be impractical to implement S3RL models on consumer devices, their knowledge can still aid in designing more compact target models. For instance, knowledge distillation (KD)~\cite{hinton_kd} allows for transferring knowledge from a larger model to a smaller one, resulting in comparable generalization capabilities. %Their main difference relies on the pretext task that each model is trying to solve. \emph{Wav2Vec 2.0} employs contrastive loss to predict quantized representations for missing slots. In contrast, WavLM employs a BERT-like classification loss over masked regions. Additionally, the WavLM training set also contains diverse datasets and data augmentation techniques to learn noise-invariant features.

% While it may be impractical to implement S3RL models on consumer devices, their knowledge can still be leveraged to aid in the design of our more compact target models. Knowledge distillation (KD), as proposed in~\cite{hinton_kd}, allows for the transfer of knowledge from a larger model to a smaller one, resulting in comparable generalization capabilities.

Moreover, it is known that edge processing can provide an extra layer of security protection for the user, ensuring that private data is processed locally on the device and not sent over the cloud to a third-party server. However, recent research has shown that KWS systems based on self-supervised representations can be vulnerable to so-called adversarial attacks during inference time \cite{paper_smc}. Adversarial attacks aim to design a small perturbation $\delta$ that when added to the test signal will force the system to fail. Over-the-air adversarial attacks, for example, have shown that imperceptible noise can be added to the user's voice resulting in misclassifications~\cite{10078009}. 

Adversarial training (AT)~\cite{madry2018towards} has emerged as a potential defense technique against adversarial attacks. AT is a robust optimization formulation that, in practice, can be seen as a data augmentation technique that generates adversarial versions of a natural sample to be used during training of Teacher models. The TRADES (TRadeoff-inspired Adversarial DEfense via Surrogate-loss minimization) method, for example, has generated inspiring results~\cite{zhang2019theoretically}. Efficiently designing small (Student) models, however, is a challenging task. The Adversarial Robust Distillation (ARD)~\cite{goldblum2020adversarially} method combines adversarial training with KD, emphasizing the importance of minimizing the Kullback-Leibler divergence between Student and Teacher logits to enhance Student robustness. In contrast, the Robust Soft Label Adversarial Distillation (RSLAD)~\cite{zi2021revisiting} method uses soft labels from the Teacher model (in lieu of hard labels) for guidance during distillation, yielding better results than ARD empirically. Notwithstanding, neither of these methods can consistently surpass the accuracy of a smaller model trained with AT, thus suggesting that further innovations are needed if large self-supervised speech representations are to be used in edge devices.

% In this work, we propose VIC-KD, a novel distillation recipe that not only compresses the model but jointly increases its robustness against adversarial attacks. In particular, we show that multi-view inputs and geometric constraints on the latent space of the student model are essential to achieve such advantages. In order to assess the benefits of our proposed method, we restrict our studies to student models with less than 96K parameters and 3 MMACs. We distill the knowledge from finetuned versions of \emph{Wav2Vec 2.0} and WavLM, showing that our distillation recipe can achieve better robustness than traditional defense techniques, such as TRADES, and improve upon robust distillation methods, such as ARD and RSLAD.

In this work, we propose VIC-KD, a novel distillation recipe that compresses the S3RL model size while increasing its robustness against adversarial attacks. In particular, we show that multi-view inputs and geometric constraints on the latent space of the Student model are essential to achieve these advantages. We consider Student models with fewer than 96K parameters and 3 MMACs, and knowledge distillation is performed from fine-tuned versions of \emph{Wav2Vec 2.0} and WavLM. Results show that our distillation recipe can achieve better robustness than traditional defense techniques, such as TRADES, and improve upon robust distillation methods, such as ARD and RSLAD.

\section{Variance-Invariance-Covariance Knowledge Distillation}
\label{sec:pagestyle}

\begin{table*}[htb]
    \centering
    \caption{Baseline results obtained directly in a supervised setting without knowledge distillation.}
    \label{tab:baseline}
    \resizebox{0.85\linewidth}{!}{%
        \begin{tabular}{ccccccccccc}
            \toprule
            & & & \multicolumn{4}{c}{Natural Training} & \multicolumn{4}{c}{Robust Training}\\
            \cmidrule(lr){4-7}
            \cmidrule(lr){8-11}
            & & & \multicolumn{2}{c}{GSC v12} & \multicolumn{2}{c}{GSC v35} & \multicolumn{2}{c}{GSC v12} & \multicolumn{2}{c}{GSC v35}\\
            \cmidrule(lr){4-5}
            \cmidrule(lr){6-7}
            \cmidrule(lr){8-9}
            \cmidrule(lr){10-11}
             Model & \#Params & MMACs & Clean & AutoAttack & Clean & AutoAttack & Clean & AutoAttack & Clean & AutoAttack \\
            \midrule
            \emph{Wav2Vec 2.0} & 94.4M & 6160.9 & 99.07 & 9.40 & 97.53 & 2.68 & 97.58 & 93.80 & 95.77 & 90.63 \\
            WavLM & 94.4M & 6132.7 & 98.94 & 6.48 & 97.42 & 0.64 & 97.95 & 94.79 & 97.20 & 93.26 \\

            \hdashline

            XVector & 80.4K & 2.9 & 96.43 & 3.09 & 94.54 & 1.36 & 94.70 & 80.33 & 93.56 & 78.27 \\
            TC-ResNet & 67.7K & 1.6 & 94.95 & 2.67 & 93.53 & 0.09 & 93.69 & 79.25 & 92.95 & 77.30 \\

            \midrule[\heavyrulewidth]
            \bottomrule
        \end{tabular} %\vspace{-6mm}
    }
\end{table*}

Herein, instead of using the Teacher model to guide how the Student logits should behave, we induce some geometric properties of the latent space of the Student via the variance-invariance-covariance regularization and the usage of multi-view inputs to each model. Although the usage of multi-view inputs for KD has already been explored in the literature to increase the environmental robustness of Student models~\cite{10022474, guimaraes2023robustdistiller}, its effect on robust distillation still needs to be determined. Figure~\ref{fig:rgkd_diagram} depicts a block diagram of the proposed VIC-KD distillation recipe.

In the self-supervised learning literature, joint embedding architectures (JEA) are becoming popular due to their effectiveness in learning latent factors from data~\cite{grill2020bootstrap, zbontar2021barlow}. VICReg~\cite{bardes2022vicreg} is one such method that is based on preserving the information of the embeddings while avoiding representational collapse. Here, we expand those ideas to a robust distillation method. First, we sample an utterance from $x \sim \mathcal{D}_{\text{train}}$ and two random transformations $\{t, t' \sim \mathcal{T} | t \neq t' \}$. As depicted in the upper branch of Figure~\ref{fig:rgkd_diagram}, the Teacher model $T_\theta$ is responsible for extracting representations from the speech input. More precisely, representation $Z$ is extracted, which is the weighted sum of the intermediate representations from all Transformer layers, and aggregated over the time dimension.

The Student branch, on the other hand, receives the other utterance view generated by $t'$ with an added adversary perturbation. First, for both inputs, we extract the latent representations of the Student model, denoted by $H'$. This representation is then fed to a classification head, generating the model logits $Y'$ that attempt to predict the spoken commands. In parallel, $H'$ is fed to a projection head responsible for generating $Z'$, which has the same dimensionality as the Teacher latent representation. Note that, at test time, only the Student encoder $S_{\theta'}$ and the classification head are used.

In addition to the mechanism described above, we now describe the contribution of each term that composes the three-factor VIC-KD loss function, presented in equation~\ref{eq:vicregloss}: (1) variance, (2) invariance, and (3) covariance terms. Following our previous notation, we define $Z = [z_1, z_2, ..., z_n]$ and $Z' = [z_1', z_2', ..., z_n']$ as the $d$-dimensional latent representations generated by the Teacher and Student models for natural and adversarially perturbed input, respectively, for a batch of $n$ utterances. First, the invariance term consists of the mean squared error between the two latent representations, $Z$ and $Z'$. Next, the variance term induces the learned latent embedding to vary within the other batch elements, thus avoiding collapse on the same vector. Lastly, the covariance term prevents the Student model, which already has limited capacity, from encoding similar features. In our experiments, we consider an equal contribution of each term in the final VICReg loss. The interested reader is referred to \cite{bardes2022vicreg} for more details about the loss implementation.
\vspace{-2mm}
\begin{align}
\mathcal{L}_{\text{VICReg}} &= \text{Var}(Z') + \text{Inv}(Z, Z') + \text{Cov}(Z')\label{eq:vicregloss}
\end{align}

%Following our previous notation, we define $Z = [z_1, z_2, ..., z_n]$ and $Z' = [z_1', z_2', ..., z_n']$ as the $d$-dimensional latent representations generated by the Teacher and Student models for natural and adversarially perturbed input, respectively, for a batch of $n$ utterances.

% \begin{align}
% \mathcal{L}_{\text{VICReg}} &= \text{Var}(Z') + \text{Inv}(Z, Z') + \text{Cov}(Z') \\
% \text{Var}(Z') &= \frac{1}{d}\sum_{k}{\text{max}\left(0, 1 - \sqrt{\text{Var}(z_k')+\epsilon}\right)} \label{eq:lvar} \\
% \text{Inv}(Z, Z') &= \frac{1}{n}\sum_{k}{||z_k - z_k'||^2_2} \label{eq:linv} \\
% \text{Cov}(Z') &= \frac{1}{d}\sum_{i \neq j}{\left[\frac{1}{n-1}\sum_{k=1}^{n}{(z_k' - \bar{z}')(z_k' - \bar{z}')^\intercal}\right]^2_{i,j}} \label{eq:lcov}
% \end{align}

% First, the invariance term consists of the mean squared error between the two latent representations, presented in equation~\ref{eq:linv}. The variance term, shown in equation~\ref{eq:lvar}, induces the learned latent embedding to vary within the other batch elements, thus avoiding collapse on the same vector. Lastly, the covariance term,  equation~\ref{eq:lcov}, prevents our student model, which already has limited capacity, from encoding similar features. In our experiments, we consider an equal contribution of each term in the final VICReg loss.

Finally, the VIC-KD loss is computed as the convex combination between the TRADES~\cite{zhang2019theoretically} and VICReg losses, controlled through a hyperparameter $\alpha$ as described in Figure~\ref{fig:rgkd_diagram}.

% To solve the inner term, the maximization problem, we usually rely on techniques to generate adversarial samples, such as the Projected Gradient Descent (PGD).

\vspace{-3mm}
\section{Experimental Setup}
\label{sec:typestyle}
\vspace{-2mm}

% \subsection{Datasets}

To train the KWS system, we use the \emph{Google Speech Commands v0.02} (GSC) dataset~\cite{warden2018speech}. This dataset has about 100k 1-second utterances spread across 35 commands and sampled at 16 kHz. We consider two versions of the dataset, one with all commands and another with only 12 classes that include labels such as \{yes, no, up, down, left, right, on, off, stop, go, unknown, and silence\}. The \emph{silence} class is made up of background noises with no speech, while the \emph{unknown} label includes utterances uniformly sampled from unused classes. We follow the SpeechBrain~\cite{speechbrain} recipe, where the dataset is split into train, validation, and test sets in the ratio of 80\%, 10\%, and 10\%, respectively.

% \subsection{Baselines and Implementation Details}

% In this study, we explore two teacher models, \emph{Wav2Vec 2.0} and the WavLM, and two student models, the TC-ResNet-8~\cite{choi19_interspeech} and the XVector~\cite{8461375}. We chose those student models considering their suitability for deployment on edge devices. For the XVector, we implemented a custom version with only 3 TDNN layers and smaller kernel sizes to fit our resource constraints. Furthermore, as summarized in table~\ref{tab:baseline}, we report the baseline results for both dataset configurations (12 or 35 classes) and show the results of clean and robust accuracy. In addition, we consider both scenarios of training the models in a standard versus an adversarial fashion via TRADES. Our following experiments will use those reference values to compare with the distillation recipes.

In this study, we examine two Teacher models, \emph{Wav2Vec 2.0} and WavLM, along with two Student models, TC-ResNet-8~\cite{choi19_interspeech} and a custom XVector~\cite{8461375}, with a reduced number of TDNN layers and smaller kernel sizes to fit our resource constraints. Furthermore, as summarized in Table~\ref{tab:baseline}, we show the results with conventional training, as well as AT via TRADES. Moreover, we report accuracy for clean speech files, as well as speech files corrupted by the AutoAttack method~\cite{croce2020reliable}. Experiments herein rely on these reference values for comparison.

Baseline models are trained for 100 epochs, with a batch size of 32 samples, using an Adam optimizer, and learning of $10^{-3}$ that linearly decays to $10^{-4}$. Conversely, we fine-tune the Teacher model for the KWS task for 10 epochs, and the learning rate is scheduled from $5\times10^{-4}$ to $5\times10^{-5}$. The Teacher model consists of the S3RL encoder with a weighted sum and aggregation mechanism described previously and a simple linear layer that classifies the input speech signal into the desired number of classes. Note that, for VIC-KD, we can discard the Teacher's linear layer since we are interested in the latent representation. For adversarial attacks, we rely on AutoAttack~\cite{croce2020reliable}, an ensemble method comprised of APGD~\cite{croce2020reliable}, APGD-T~\cite{croce2020reliable}, and FAB~\cite{croce2020minimally} attacks. We consider $\ell_\infty$ bounded attacks with an $\epsilon = 1.5\times10^{-3}$.

% We selected three recipes to compare the distillation methods against our proposed technique: KD, ARD, and RSLAD. We follow the recipes from their respective papers as closely as possible. In particular, when adapting to our KWS use case, we train the models for $250$ epochs and the same parameters as the small baseline models, as specified above. For the VIC-KD adversarial training, we implement a $10$-steps PGD-like attack to generate the perturbation, also with an $\epsilon = 1.5\times10^{-3}$ and a step size of $3\times10^{-4}$. Besides the robust training, our distillation recipe also proposes using multi-view inputs to student and teacher models. Six key elements in the set $\mathcal{T}$ of transformations are used: using clean data, adding noise, reverberation, and noise-plus-reverberation, dropping wave chunks, and speed perturbation.

Three distillation methods, namely KD, ARD, and RSLAD, are used to benchmark the proposed technique. Training is performed for 250 epochs with parameters matching the small baseline models, as specified above. VIC-KD relies on a 10-step PGD-like attack, also with an $\epsilon = 1.5\times10^{-3}$ and a step size of $3\times10^{-4}$, to generate the adversarial perturbation for training. Additionally, multi-view inputs are used for Student and Teacher models, incorporating clean data, noise, reverberation, noise-plus-reverberation, wave chunk dropping, and speed perturbation into the transformations $\mathcal{T}$.

\section{Experimental Results and Discussion}
\label{sec:majhead}

\subsection{Classification accuracy}

% \vspace{-4mm}
% Vou remover os valores percentuais do KD
% ~{\small \color{gray}(---)}
\begin{table}[htb]
    \centering
    \caption{Results on the GSC dataset with 12 classes for distillation methods. Distilling from standard and robust Teachers.}
    \label{tab:main_results}
    \resizebox{\linewidth}{!}{%
        \begin{tabular}{ccccc}
                \toprule
                \multirow{2}{*}{Teacher / Student} & \multicolumn{2}{c}{Standard Teacher} & \multicolumn{2}{c}{Robust Teacher} \\
                \cmidrule(lr){2-3}
                \cmidrule(lr){4-5}

                 & Clean & AutoAttack & Clean & AutoAttack \\
                \midrule
                \multicolumn{5}{c}{KD~\cite{hinton_kd}} \\
                \midrule
                \emph{Wav2Vec 2.0} / TC-ResNet & 95.80 & 5.54 & 95.61 & 4.49 \\ %~{\small \color{red}(-74.8\%)} \\
                \emph{Wav2Vec 2.0} / XVector & 96.77 & 7.59 & 96.68 & 6.84 \\ %~{\small \color{red}(-73.5\%)} \\
                WavLM / TC-ResNet & 96.20 & 9.29 & 95.70 & 5.11 \\ %~{\small \color{red}(-74.1\%)} \\
                WavLM / XVector & \textbf{96.91} & 8.00 & \textbf{96.62} & 7.19 \\ %~{\small \color{red}(-73.1\%)} \\

                \midrule
                \multicolumn{5}{c}{ARD~\cite{goldblum2020adversarially}} \\
                \midrule
                \emph{Wav2Vec 2.0} / TC-ResNet & 94.54 & 76.41~{\small \color{red}(-2.8\%)} & 94.92 & 74.16~{\small \color{red}(-5.1\%)} \\
                \emph{Wav2Vec 2.0} / XVector & 95.37 & 79.74~{\small \color{red}(-0.6\%)} & 95.71 & 79.94~{\small \color{red}(-0.4\%)} \\
                WavLM / TC-ResNet & 95.17 & 74.13~{\small \color{red}(-5.1\%)} & 95.30 & 76.87~{\small \color{red}(-2.4\%)} \\
                WavLM / XVector & 95.27 & 78.33~{\small \color{red}(-2.0\%)} & 95.29 & 74.03~{\small \color{red}(-6.3\%)} \\

                \midrule
                \multicolumn{5}{c}{RSLAD~\cite{zi2021revisiting}} \\
                \midrule
                \emph{Wav2Vec 2.0} / TC-ResNet & 94.65 & 78.06~{\small \color{red}(-1.2\%)} & 94.59 & 76.27~{\small \color{red}(-3.0\%)} \\
                \emph{Wav2Vec 2.0} / XVector & 95.37 & 80.79~{\small \color{Green}(+0.5\%)} & 95.45 & 81.98~{\small \color{Green}(+1.7\%)} \\
                WavLM / TC-ResNet & 95.11 & 77.17~{\small \color{red}(-2.1\%)} & 94.11 & 77.99~{\small \color{red}(-1.3\%)} \\
                WavLM / XVector & 95.34 & 81.30~{\small \color{Green}(+1.0\%)} & 95.69 & 81.90~{\small \color{Green}(+1.6\%)} \\

                \midrule
                \multicolumn{5}{c}{VIC-KD (\textbf{Ours})} \\
                \midrule
                \emph{Wav2Vec 2.0} / TC-ResNet & 95.31 & 83.75~{\small \color{Green}(+4.5\%)} & 95.48 & 83.38~{\small \color{Green}(+4.1\%)} \\
                \emph{Wav2Vec 2.0} / XVector & \cellcolor{gray!25}\textbf{96.50} & \cellcolor{gray!25}\textbf{86.12~{\small \color{Green}(+5.8\%)}} & 96.29 & 85.33~{\small \color{Green}(+5.0\%)} \\
                WavLM / TC-ResNet & 95.39 & 83.30~{\small \color{Green}(+4.1\%)} & 95.37 & 83.73~{\small \color{Green}(+4.5\%)} \\ 
                WavLM / XVector & 95.92 & 86.08~{\small \color{Green}(+5.8\%)} & \cellcolor{gray!25}\textbf{96.39} & \cellcolor{gray!25}\textbf{86.31~{\small \color{Green}(+6.0\%)}} \\
                \midrule[\heavyrulewidth]
                \bottomrule
            \end{tabular} %\vspace{-6mm}
    }
\end{table}

Table~\ref{tab:main_results} presents the main experimental results exploring two scenarios: standard vs. robust Teachers as guides. We report clean and robust accuracies for each robust distillation method, with differences relative to the respective baseline in parentheses. In fact, given the low robust accuracy and that KD is not a robust method, we do not compute its delta against the robust baseline. KD with standard Teacher outperforms baselines and other robust methods in clean accuracy. However, despite some improvement, the robust accuracy is still below random guess. ARD, on the other hand, significantly enhances overall robust accuracy, but still falls short of the baseline performance under the robust condition.

Regardless, we observe an improvement in the overall scenario for RSLAD and VIC-KD, which are built upon TRADES. For the Student model with bigger capacity, e.g., the XVector, we observe the first results where the model can improve upon the baseline for robust accuracy. The TC-ResNet on the RSLAD distillation has inline results with the baseline, but improves upon the ARD method. On the other hand, VIC-KD for both TC-ResNet and XVector can substantially improve upon both robust distillation recipes and the baselines for the clean and under-attack scenarios, thus showing the benefits of the proposed methodology. In fact, VIC-KD with the \emph{Wav2Vec 2.0} / XVector pair outperforms the baseline by a relative percentual difference of 7.2\% on robust accuracy. Similarly, the proposed model outperforms ARD and RSLAD on attacked scenarios by 8.0\% and 6.6\%, respectively. On the downside, the VIC-KD recipe training time is four times slower than RSLAD. Note that, at the inference stage, the time is the same for all distillation recipes since it depends solely on the architecture of the student model.

Finally, some conclusions can be drawn from the distillation of robust Teachers. First, from the KD experiment, it is crucial to notice that the Student model does not necessarily inherit robustness from the Teacher; hence, specific techniques are needed if the goal is to design adversarially robust models. However, a tradeoff between clean and robust accuracy needs to be decided. For the robust distillation methods, we observe that using robust Teachers, in general, does not improve the final performance of the Student model; thus, for our specific use case on the GSC dataset, unless one already has off-the-shelf robust Teacher models, it is not worth the extra computation of transforming a standard Teacher into a robust one first and then perform the robust distillation.

\subsection{The effects of multi-view inputs}
\begin{figure}[htb!]
        \centering
        \includegraphics[width=\linewidth]{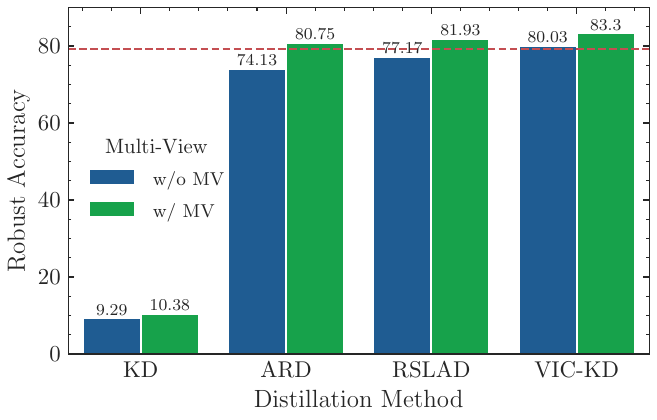} \vspace{-8mm}
        \caption{Effects of the Multi-View on several distillation Methods for the WavLM / TC-ResNet pair. The dashed line represents the robust accuracy of TC-ResNet via TRADES.}
        \label{fig:effect_mv} \vspace{-2mm}
\end{figure}

Next, we investigate the effect of multi-view (MV) inputs on robust distillation, as shown in Fig.\ref{fig:effect_mv}. Here, we investigate the WavLM/TC-ResNet pair as our Teacher and Student models, respectively. For comparison, a red dashed line shows the baseline robust accuracy of TC-ResNet trained with TRADES without any guidance from Teacher models. As observed in the figure, all distillation methods benefit from multi-view. For instance, ARD and RSLAD, which had a robust accuracy below the baseline, can surpass this landmark after MV. Our proposed VIC-KD method can outperform the baseline with or without MV. However, MV helps us to achieve the best overall robust accuracy.

We hypothesize that the reason behind the performance gain by using MV is two-fold. First, MV induces the student model to learn perturbation-invariant features from speech signals and to better disentangle noisy factors from speech features, thus improving generalization. Lastly, other works already have discussed that self-supervised learning methods that employ MV are forcing the model to maximize the mutual information of representations between the two views~\cite{bardes2022vicreg}. We suppose a similar conclusion can be drawn for the KD case, but more studies are still needed.

\subsection{Expanding the distillation to more classes}

\begin{figure}[htb]
        \centering
        \includegraphics[width=0.8\linewidth]{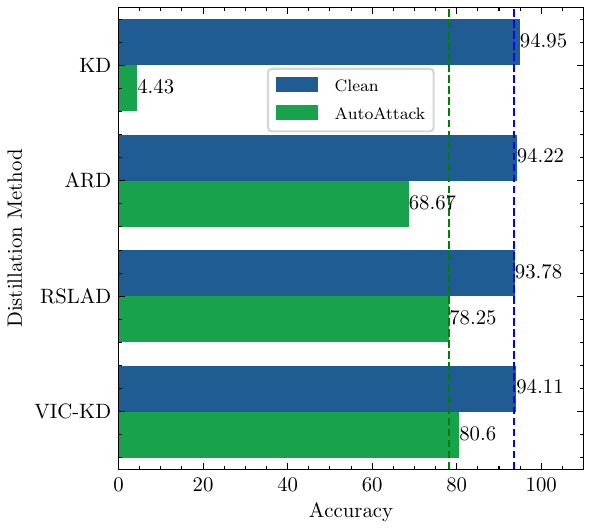} \vspace{-5mm}
        \caption{Scaling robust distillation to a 35-class GSC dataset and comparing it to clean and robust accuracy baselines.}
        \label{fig:gsc35} \vspace{-2mm}
\end{figure}

Lastly, we further stress the distillation recipes to a more significant number of output classes by using all 35 classes from the GSC dataset. In Fig.~\ref{fig:gsc35}, we show the accuracy results for both clean and AutoAttack-based methods using the WavLM / XVector models as Teacher and Student respectively. The blue and green dashed lines denote the clean and robust accuracy of the supervised baseline, XVector, trained via TRADES. Similarly to previous findings, KD can surpass the clean accuracy of the baseline, but the robust accuracy is not satisfactory. Among the robust distillation methods, VIC-KD exhibits the overall best performance by achieving the best accuracy and showing the smallest gap between clean and robust accuracies.

\vspace{-2.5mm}
\section{Conclusions}
\label{sec:conclusion}
\vspace{-2mm}
Here, we propose VIC-KD, a methodology to improve the adversarial robustness of distilled models. Our recipe is two-fold: (1) use multi-view inputs to induce the Student to learn perturbation-invariant features and (2) apply the variance-invariance-covariance regularization (VICReg) to the latent representation of the Teacher/Student model. Experiments on the Google Speech Command dataset, with 12 and 35 classes, show the proposed methodology outperforming state-of-the-art robust distillation recipes on $\ell_\infty$-bounded ensemble of attack (AutoAttack). Overall, VIC-KD canbetter balance the tradeoff between clean and robust accuracy, making this technique a strong candidate for developing and deploying trustworthy speech applications on the edge.

% Below is an example of how to insert images. Delete the ``\vspace'' line,
% uncomment the preceding line ``\centerline...'' and replace ``imageX.ps''
% with a suitable PostScript file name.
% -------------------------------------------------------------------------
% \begin{figure}[htb]

% \begin{minipage}[b]{1.0\linewidth}
%   \centering
%   \centerline{\includegraphics[width=8.5cm]{image1}}
% %  \vspace{2.0cm}
%   \centerline{(a) Result 1}\medskip
% \end{minipage}
% %
% \begin{minipage}[b]{.48\linewidth}
%   \centering
%   \centerline{\includegraphics[width=4.0cm]{image3}}
% %  \vspace{1.5cm}
%   \centerline{(b) Results 3}\medskip
% \end{minipage}
% \hfill
% \begin{minipage}[b]{0.48\linewidth}
%   \centering
%   \centerline{\includegraphics[width=4.0cm]{image4}}
% %  \vspace{1.5cm}
%   \centerline{(c) Result 4}\medskip
% \end{minipage}
% %
% \caption{Example of placing a figure with experimental results.}
% \label{fig:res}
% %
% \end{figure}

\bibliographystyle{IEEEbib}
\bibliography{strings,refs}

\begin{thebibliography}{10}

\bibitem{chen2014small}
G.~Chen, C.~Parada, and G.~Heigold,
\newblock ``Small-footprint keyword spotting using deep neural networks,''
\newblock in {\em 2014 IEEE international conference on acoustics, speech and
  signal processing (ICASSP)}, 2014, pp. 4087--4091.

\bibitem{OSHAUGHNESSY2023101538}
D.~O'Shaughnessy,
\newblock ``Trends and developments in automatic speech recognition research,''
\newblock {\em Computer Speech \& Language}, vol. 83, pp. 101538, 2023.

\bibitem{8388669}
A.~R. Avila, J.~Monteiro, D.~O'Shaughneussy, and T.~H. Falk,
\newblock ``Speech emotion recognition on mobile devices based on modulation
  spectral feature pooling and deep neural networks,''
\newblock in {\em 2017 IEEE International Symposium on Signal Processing and
  Information Technology (ISSPIT)}, 2017, pp. 360--365.

\bibitem{baevski2020wav2vec}
A.~Baevski, Y.~Zhou, A.~Mohamed, et~al.,
\newblock ``wav2vec 2.0: A framework for self-supervised learning of speech
  representations,''
\newblock {\em Advances in Neural Information Processing Systems}, vol. 33, pp.
  12449--12460, 2020.

\bibitem{chen2022wavlm}
S.~Chen, C.~Wang, Z.~Chen, et~al.,
\newblock ``Wavlm: Large-scale self-supervised pre-training for full stack
  speech processing,''
\newblock {\em IEEE Journal of Selected Topics in Signal Processing}, 2022.

\bibitem{hinton_kd}
G.~Hinton, O.~Vinyals, and J.~Dean,
\newblock ``Distilling the knowledge in a neural network,''
\newblock in {\em NIPS Deep Learning and Representation Learning Workshop},
  2015.

\bibitem{paper_smc}
H.~R. Guimarães, Y.~Zhu, O.~Mengara, et~al.,
\newblock ``Assessing the vulnerability of self-supervised speech
  representations for keyword spotting under white-box adversarial attacks,''
\newblock in {\em 2023 IEEE International Conference on Systems, Man, and
  Cybernetics (SMC)}, 2023.

\bibitem{10078009}
C.~Zhao, Z.~Li, H.~Ding, et~al.,
\newblock ``Utio: Universal, targeted, imperceptible and over-the-air audio
  adversarial example,''
\newblock in {\em 2022 IEEE 28th International Conference on Parallel and
  Distributed Systems}, 2023, pp. 346--353.

\bibitem{madry2018towards}
A.~Madry, A.~Makelov, L.~Schmidt, et~al.,
\newblock ``Towards deep learning models resistant to adversarial attacks,''
\newblock in {\em International Conference on Learning Representations}, 2018.

\bibitem{zhang2019theoretically}
H.~Zhang, Y.~Yu, J.~Jiao, E.~Xing, et~al.,
\newblock ``Theoretically principled trade-off between robustness and
  accuracy,''
\newblock in {\em International conference on machine learning}. PMLR, 2019,
  pp. 7472--7482.

\bibitem{goldblum2020adversarially}
M.~Goldblum, L.~Fowl, S.~Feizi, et~al.,
\newblock ``Adversarially robust distillation,''
\newblock in {\em Proc. of the AAAI Conference on Artificial Intelligence},
  2020, vol.~34, pp. 3996--4003.

\bibitem{zi2021revisiting}
B.~Zi, S.~Zhao, X.~Ma, et~al.,
\newblock ``Revisiting adversarial robustness distillation: Robust soft labels
  make student better,''
\newblock in {\em Proceedings of the IEEE/CVF International Conference on
  Computer Vision}, 2021, pp. 16443--16452.

\bibitem{10022474}
K.P. Huang, Y.K. Fu, T.Y. Hsu, et~al.,
\newblock ``Improving generalizability of distilled self-supervised speech
  processing models under distorted settings,''
\newblock in {\em 2022 IEEE Spoken Language Technology Workshop}, 2023, pp.
  1112--1119.

\bibitem{guimaraes2023robustdistiller}
H.~R. Guimarães, A.~Pimentel, A.~R. Avila, et~al.,
\newblock ``Robustdistiller: Compressing universal speech representations for
  enhanced environment robustness,''
\newblock in {\em 2023 IEEE International Conference on Acoustics, Speech and
  Signal Processing (ICASSP)}, 2023, pp. 1--5.

\bibitem{grill2020bootstrap}
J.B. Grill, F.~Strub, F.~Altch{\'e}, et~al.,
\newblock ``Bootstrap your own latent-a new approach to self-supervised
  learning,''
\newblock {\em Advances in neural information processing systems}, vol. 33, pp.
  21271--21284, 2020.

\bibitem{zbontar2021barlow}
J.~Zbontar, Li~Jing, I.~Misra, et~al.,
\newblock ``Barlow twins: Self-supervised learning via redundancy reduction,''
\newblock in {\em International Conference on Machine Learning}. PMLR, 2021,
  pp. 12310--12320.

\bibitem{bardes2022vicreg}
A.~Bardes, J.~Ponce, and Y.~LeCun,
\newblock ``{VICR}eg: Variance-invariance-covariance regularization for
  self-supervised learning,''
\newblock in {\em International Conference on Learning Representations}, 2022.

\bibitem{warden2018speech}
P.~Warden,
\newblock ``Speech commands: A dataset for limited-vocabulary speech
  recognition,''
\newblock {\em arXiv preprint arXiv:1804.03209}, 2018.

\bibitem{speechbrain}
M.~Ravanelli, T.~Parcollet, P.~Plantinga, et~al.,
\newblock ``{SpeechBrain}: A general-purpose speech toolkit,'' 2021,
\newblock arXiv:2106.04624.

\bibitem{choi19_interspeech}
S.~Choi, S.~Seo, B.~Shin, et~al.,
\newblock ``{Temporal Convolution for Real-Time Keyword Spotting on Mobile
  Devices},''
\newblock in {\em Proc. Interspeech 2019}, 2019, pp. 3372--3376.

\bibitem{8461375}
D.~Snyder, D.~Garcia-Romero, G.~Sell, et~al.,
\newblock ``X-vectors: Robust dnn embeddings for speaker recognition,''
\newblock in {\em 2018 IEEE International Conference on Acoustics, Speech and
  Signal Processing (ICASSP)}, 2018, pp. 5329--5333.

\bibitem{croce2020reliable}
F.~Croce and M.~Hein,
\newblock ``Reliable evaluation of adversarial robustness with an ensemble of
  diverse parameter-free attacks,''
\newblock in {\em International conference on machine learning}. PMLR, 2020,
  pp. 2206--2216.

\bibitem{croce2020minimally}
F.~Croce and M.~Hein,
\newblock ``Minimally distorted adversarial examples with a fast adaptive
  boundary attack,''
\newblock in {\em International Conference on Machine Learning}. PMLR, 2020,
  pp. 2196--2205.

\end{thebibliography}

\end{document}